\documentclass[reprint,amsmath,amssymb,aps,superscriptaddress]{revtex4-2}

\usepackage{braket}
\usepackage{graphicx}
\usepackage{subcaption}
\usepackage{caption}
\usepackage{graphicx}% Include figure files
\usepackage{dcolumn}% Align table columns on decimal point
\usepackage{bm}% bold math
\usepackage{hyperref}
\hypersetup{colorlinks, citecolor=blue, linkcolor=blue, urlcolor=blue}
\usepackage{comment}
\newcommand{\req}[1]{Eq.~(\ref{#1})}
\newcommand{\kB}{k_\textrm{B}}
\newcommand{\rfig}[1]{Fig.~\ref{#1}}
\newcommand{\rsec}[1]{Section~\ref{#1}}
\newcommand{\rtab}[1]{Table~\ref{#1}}
\newcommand{\rapp}[1]{Appendix~\ref{#1}}
\newcommand{\murm}{%
  \ifmmode
    \mathchoice
        {\hbox{\normalsize\textmu}}
        {\hbox{\normalsize\textmu}}
        {\hbox{\scriptsize\textmu}}
        {\hbox{\tiny\textmu}}%
  \else
    \textmu
  \fi
}
\newcommand{\micrometer}{\murm{\rm m}\xspace}
\newcommand{\microsecond}{\murm{\rm s}\xspace}

\usepackage{tikz}
\usetikzlibrary{arrows,shapes,snakes,automata,backgrounds,petri,calc}
\usetikzlibrary{decorations.markings}

\usepackage[americaninductors]{circuitikz}

\usepackage{xspace}

\begin{document}

\title{Feasibility study of quantum computing using trapped electrons}% 
\author{Qian~Yu}
\email{qian$\_$yu@berkeley.edu}
\affiliation{Physics Department, University of California, Berkeley, CA 94720, (USA)}
\affiliation{Challenge Institute for Quantum Computation, University of California, Berkeley, CA 94720, (USA)}%Lines break automatically or can be forced with \\

\author{Alberto~M.~Alonso}%
\affiliation{Physics Department, University of California, Berkeley, CA 94720, (USA)}
\affiliation{Challenge Institute for Quantum Computation, University of California, Berkeley, CA 94720, (USA)}

\author{Jackie~Caminiti}%
\affiliation{Physics Department, University of California, Berkeley, CA 94720, (USA)}
\affiliation{Challenge Institute for Quantum Computation, University of California, Berkeley, CA 94720, (USA)}

\author{Kristin~M.~Beck}%
\affiliation{Lawrence Livermore National Laboratory, 7000 East Avenue Livermore, CA 94550, (USA)}

\author{R.~Tyler~Sutherland}%
\affiliation{Department of Electrical and Computer Engineering, Department of Physics and Astronomy, University of Texas at San Antonio, San Antonio, TX 78249, (USA)}

\author{Dietrich~Leibfried}%
\affiliation{Time and Frequency Division, National Institute of Standards and Technology, 325 Broadway, Boulder, CO 80305, (USA)}

\author{Kayla~J.~Rodriguez}%
\affiliation{Physics Department, University of California, Riverside, CA 92521, (USA)}

\author{Madhav~Dhital}%
\affiliation{Physics Department, University of California, Riverside, CA 92521, (USA)}

\author{Boerge~Hemmerling}%
\affiliation{Physics Department, University of California, Riverside, CA 92521, (USA)}

\author{Hartmut~H\"affner}%
\affiliation{Physics Department, University of California, Berkeley, CA 94720, (USA)}
\affiliation{Challenge Institute for Quantum Computation, University of California, Berkeley, CA 94720, (USA)}
\affiliation{Computational Research Division, Lawrence Berkeley National Laboratory, Berkeley, CA 94720, (USA)}

\date{\today}

\begin{abstract}
We investigate the feasibility of using electrons in a linear Paul trap as qubits in a future quantum computer. We discuss the necessary experimental steps to realize such a device through a concrete design proposal, including trapping, cooling, electronic detection, spin readout and single and multi-qubit gate operations. Numeric simulations indicate that two-qubit Bell-state fidelities of order 99.99\% can be achieved assuming reasonable experimental parameters. 
\end{abstract}

\maketitle

% \tableofcontents

\section{Introduction}
\label{sec:introduction}

Ongoing efforts to build a quantum computer are based on various physical implementations. One of the most established implementations is based on trapped ions in Paul traps, where qubits are encoded in the internal states of the ions' valence electrons and entangled using spin-dependent forces that couple the ions' internal states to their collective motion \cite{Molmer1999}. Trapped ions are advantageous because they exhibit coherence times exceeding 10 minutes \cite{Bollinger1991,Fisk1997,Wang2021single} and flexible connectivity \cite{Kielpinski2002CCD, Wineland1998}. Also, errors per gate as low as $10^{-6}$ \cite{Harty2014,Harty2016}, for single-qubit gates, and $10^{-3}$ \cite{Ballance2016,Gaebler2016,Srinivas2021}, for multi-qubit gates, have been achieved. However, multi-qubit operations between ions are typically relatively slow ($\sim 10$\,\microsecond) compared to, for example, superconducting qubits ($\sim 10$\,ns). In addition, the optical technology required for cooling, preparation, readout and controlling thousands of trapped ion qubits is still in its infancy \cite{Mehta2020b,Niffenegger2020b,Brown2021-materials-challenges}.

Here, we conduct a feasibility study of trapped-electron based quantum computing. Electrons are attractive for quantum computing because they are extremely light, and a natural two-level spin system (qubit) that has a large enough magnetic moment to be manipulated with well-established microwave technology and thermal reservoirs, eliminating the need for qubit control optics. The mass reduction of four orders of magnitude, relative to trapped ions, increases the motional frequencies of the particle in the trapping potential, thereby facilitating the speed for multi-qubit operations and transport.
Additionally, the electrons' two-level spin structure removes certain complications of traditional atomic and solid-state qubits such as leakage of quantum information from the computational subspace, potentially making high-fidelity operations easier and simplifying quantum error correction~\cite{Brown2018}. 
The electrons' spin degree of freedom can be initialized, coherently controlled, and measured using GHz electronics, microwaves, and a low temperature (0.4~K) reservoir, making all-electronic control feasible and thus replacing some of the optical engineering challenges required to build large scale quantum information processing devices with trapped ions.

Quantum information processing using electrons in Penning traps has been considered before~\cite{Bushev2008,Marzoli2008,Goldman2010}. In addition, there are efforts under way to use electrons trapped on superfluid films of helium and
neon~\cite{Lyon2006,Yang2016,Koolstra2019b,Kawakami2019,Zhou2021electron}, where single electrons have been detected and manipulated \cite{Koolstra2019a}.
%, however, coherent control over a single electron qubit has yet to be demonstrated. 
% Need to fix Koolstra2009(a) reference
%However, the former approach was hindered by trap anharmonicities \cite{Goldman2010}, while for the latter approach the embedding of the electrons in the superfluid film  micrometer~\cite{Koolstra2019a} may pose significant challenges which are yet undefined(?).
% current state of electron work
Quantum computing with electrons in Paul traps has been proposed and discussed in Refs.~\cite{Daniilidis2013electron,Peng2017,Kotler2017}. The first step towards this goal has been taken by trapping and detecting electrons in a room-temperature Paul trap~\cite{Matthiesen2021-electron-trapping}. Similar efforts are under way in the Noguchi group at University of Tokyo.

%Further challenges are cooling, spin direction detection and implementing multi-qubit gates.
Here we discuss basic elements for an electron quantum computer. We study the anticipated challenges of cooling electrons, detecting their spins and implementing multi-qubit gates. Furthermore, we carry out simulations of quantum gates to determine dominant error sources.

%%%%%%%%%%%%%%%%%%%%%%%%%%%%%%%%%%%%%%

\section{Prototype}
\label{sec:prototype}

\begin{figure*}
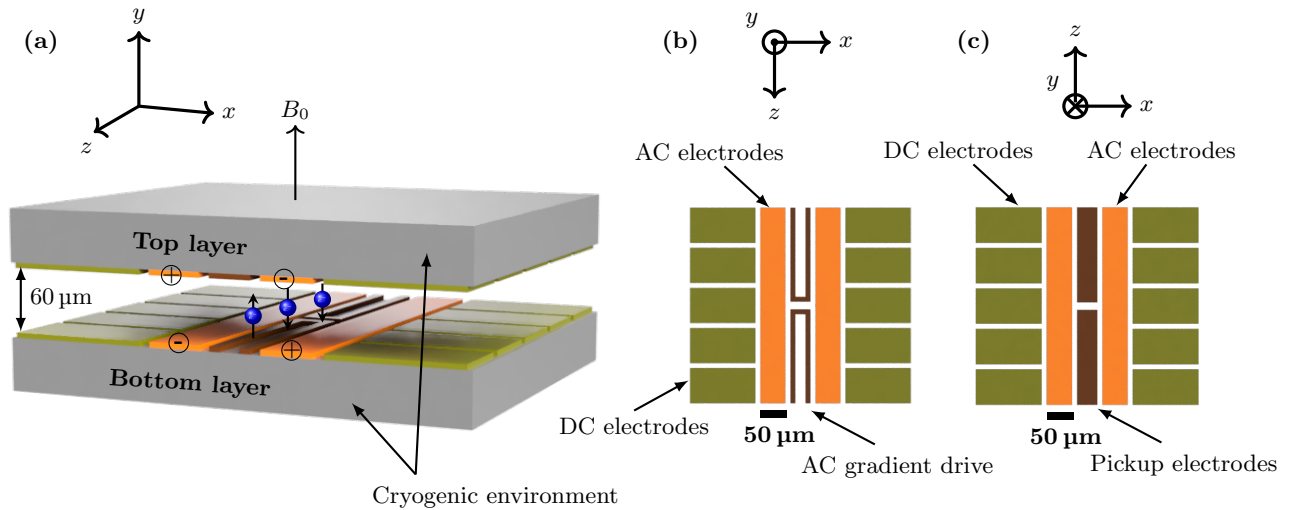


    \centering
    
    \include{fig1}
    
    \caption{
    \textbf{(a)} A prototype trap design provides a framework for analyzing the feasibility of trapped electron quantum computing. The trap consists of two substrates separated by 60~\micrometer, supporting four AC electrodes that are driven out-of-phase. Driving currents through the two center electrodes of the bottom layer generates a magnetic field gradient or a homogeneous magnetic field (both along the $z$-axis) depending on the relative phase of the currents. The two center electrodes of the top layer serve to pick up the image current generated by the electron motion. The plus and minus sign at the AC electrodes indicate the phase of the AC drive applied to each electrode. 
    \textbf{(b)} Top view of the bottom substrate. \textbf{(c)} Bottom view of the top substrate.
    \label{fig:prelimanary-trap-design-2L}
    }
\end{figure*}

%Other options for creating electrons including ....

The prototype system we consider is sketched in Fig.~\ref{fig:prelimanary-trap-design-2L} and assumed to reside in a cryogenic environment at 4\,K. We chose a quadrupole configuration consisting of two layers separated by 60~\textmu{}m. The approximately quadrupolar geometry, with a width of the quadrupole electrodes of 50~\micrometer, a length of the quadrupole electrodes of 390~\micrometer and electrode separation of 60~\micrometer, is efficient in generating both deep and stiff traps, and the symmetry suppresses odd-orders of anharmonic terms of the trapping potential.
% trapping potential
As indicated in Fig.~\ref{fig:prelimanary-trap-design-2L}, we assume to drive the two AC electrodes in each plane out of phase with each other and opposite to the other plane with an amplitude of $U_0 = 14$\,V with a frequency of $\omega_{\rm ac} = 2\pi \times 10.6$\,GHz. This yields transverse secular frequencies $\omega_{\rm t}=2\pi \times$\,2\,GHz. Further, we assume an axial trap frequency of $\omega_{\rm a} = 2\pi \times 300$\,MHz, produced by appropriate static (-15\,V to 15\,V) potentials applied to the DC electrodes shown in Fig.~\ref{fig:prelimanary-trap-design-2L}.
Using the pseudo potential approximation we find a trap depth of approximately $80$\,meV while numerical simulation of electron trajectories shows substantial losses for electrons with an initial energy above $22$\,meV (see \rapp{appendix: trap stability}).
The minimum of the trapping potential resides in the center between layers and in the middle of the electrode structure along $x$ and $z$-axis.

Low-energy electrons can be introduced into the trap by photoionizing a thermal atomic beam near the trap center \cite{Matthiesen2021-electron-trapping}. The electron motion can be damped to reliably form a Coulomb crystal and detected by coupling the induced image current to a resonant cryogenic tank circuit held at 0.4~K, which creates a voltage opposing the electron motion that is  dissipated over the effective resistance of the circuit until the thermal energy of the electron is equivalent to the circuit temperature. The average micro-motion amplitude of a thermalized electron at $T=0.4\, \rm K$ is $x_{\rm MM} = \frac{q}{2} x_{\rm t} = 80$\,nm, where $q=0.53$ is the stability parameter in the Matthieu equation, and $x_{\rm t} = \sqrt{\frac{2 k_{\rm B} T}{m_e \omega_t^2}} \approx 0.3\,$\micrometer is the average amplitude of the transverse secular motion. Here, $k_{\rm B}$ is the Boltzmann constant and $m_e$ is the electron mass. 
% detection

Quantum information will be stored in superpositions of the spin states of the electron, denoted as  $\ket{\uparrow}$ and $\ket{\downarrow}$. We assume that a homogeneous magnetic field $B_0=3.6$\,mT along the $y$-direction splits the degeneracy of the two spin states, leading to a frequency difference of $\omega_{\rm qubit}=2\pi\times 100$\,MHz between both logical eigenstates. 
The electron qubits will be manipulated with the help of the  two ``hairpin" electrodes between the AC electrodes of the bottom layer, labeled  ``AC gradient drive" in Fig.~\ref{fig:prelimanary-trap-design-2L}. 
The spin direction of the electrons is coupled to the electron motion by using an oscillating magnetic field gradient  resonant with the electron motion to apply a spin-dependent displacement force.  Measuring the phase of the resulting image current in one of the electrodes projects the spin state into one of the qubit eigenstates, as discussed in Ref.~\cite{Peng2017}. 
The oscillating magnetic field gradient is created by  currents of opposite direction in the middle segments of the hairpin electrodes and can also be used to drive multi-qubit gates.
Reversing the current direction in one of the hairpin electrodes produces an oscillating homogeneous magnetic field perpendicular to the quantization axis that can drive single-qubit gates.

\subsection{Cooling of electrons}
\label{sec:cooling}

% cooling
%Trapping the electron as close as 30~$\mathrm{\mu m}$ to a trap electrode will increase the size of the image current such that the electron will be cooled with time constants of order 2~$\mathrm{\mu s}$.

\begin{figure}
\begin{tikzpicture}

% labels
% \node at (-4.25,+3.5) {{\bf (a)}};

% \node at (3.75,3.5) {{\bf (b)}};

% \begin{scope}[shift={(-4,0.25)}]
\begin{scope}
\draw (0,0) -- (0,0.6);
\filldraw[fill=yellow!60!white, draw=black] (-0.8,0.6) rectangle (0.8,0.8);
\filldraw[fill=yellow!60!white, draw=black] (-0.8,1.7) rectangle (0.8,1.9);
\draw (0,1.9) -- (0,2.5);
\draw (0,2.5)--(0.5,2.5);
\draw (0.6,2.5) circle (0.1cm);
\node at (0.6,2.5) {{\bf -}};
\draw (-0.4,1.8) circle (0.1cm);
\draw (0,1.8) circle (0.1cm);
\draw (0.4,1.8) circle (0.1cm);
\node at (-0.4,1.8) {{\bf -}};
\node at (0,1.8) {{\bf -}};
\node at (0.4,1.8) {{\bf -}};
\draw(0.7,2.5) to [short,f=$I$,inner sep = 5pt] (1.5,2.5);
\draw (1.5,2.5) -- (3.0,2.5);
\draw (0,0) -- (3.0,0);
\draw (3.0,0) -- (3.0,0.35);
\filldraw[fill=red!40!white, draw=black] (2.75,0.35) rectangle (3.25,2.15);
\draw (3.0,2.15)--(3.0,2.5);
\node at (3,2.15) [ rotate=270, right] {$Z(\omega)$,\,0.4\,K};
% \node at (3.2,1.25) [ right] {0.4\,K};
\filldraw[fill=blue!70!white] (0,1.35) circle (0.13cm);

\draw(0,1.22) to [short,f=$v$,inner sep = 5pt] (0,1.219);
\draw (-0.4,1.25) [rotate=270] parabola (-0.75,1.75);
\draw (-0.4,1.25) [rotate=270] parabola (-0.05,1.75);
\draw (0,0.95) to (0,1.55);
\draw (-0.15,1) to (-0.15,1.5);
\draw (-0.3,1.1) to (-0.3,1.4);
\shade[ball color=blue] (0,1.35) circle (.13cm);
\node at (0,1.35) [white,scale=0.85] {$e$};
\draw
    % (3.0,2.0) circle (0.05cm)
    (3.0,2.3) to [short,-*] (4.0,2.3)
    % (3.0,0.5) circle (0.05cm)
    (3.0,0.2) to [short,-*] (4.0,0.2)
    (4.0,2.3) to [open, v^>=$U$] (4.0,0.2)
  
    ;

% \draw [arrow] (0.2,2.7) -- node[anchor=south] {I} (0.7,2.7);
% \draw

%     (0,0) to (0,1.5)
%     (0,1.5) to [american inductor, inductors/scale=1.0, inductors/width=0.5, inductors/coils=5,l_=$L_e$] (0,2)
%     (0,2) to [short] (0,2.5)
%     (0,2.5) to [short,f=$I$,inner sep = 5pt] (3,2.5)
%     (0,0) to [short] (3,0)
%     (3,0) to [short,-*] (3,0.5)
%     (3,2.5) to [short,-*] (3,2.0)

%     (3-0.5,2.0) to [short] (3+0.5,2.0)
%     (3-0.5,0.5) to [short] (3+0.5,.5)

%     (2.5,0.5) to [C, l=$C$] (2.5,2.0)
%     (3.5,0.5) to [L, l_=$L$] (3.5,2.0)
    
%     %(3.0,2.25) to [short,*-] (4.0,2.25) 
%     %to [short] (4.0,2.0) node[rground] {}

%     %(3.0,1.75) to [short,*-] (4.0,1.75) 
%     %to [short] (4.0,2.0) node[rground] {}
    
%     (3.0,2.25) to [short,-*] (4.0,2.25)
%     (3.0,0.25) to [short,-*] (4.0,0.25)
    
%     (4.0,2.4) to [open, v^>=$U$] (4.0,0.1)
  
%     ;

\end{scope}

% \begin{scope}[shift={(4,0.25)}]

% %\draw (4.5,0.0) node[plain amp] (opamp) {Amp};

% %\ctikzset{inductor=american}
% \draw

%     (0,0) to [C,l_=$C_e$] (0,1)
%     (0,1) to [american inductor, inductors/scale=1.0, inductors/width=0.5, inductors/coils=5,l_=$L_e$] (0,2)
%     (0,2) to [short] (0,2.5)
%     (0,2.5) to [short,f=$I$,inner sep = 5pt] (3,2.5)
%     (0,0) to [short] (3,0)
%     (3,0) to [short,-*] (3,0.5)
%     (3,2.5) to [short,-*] (3,2.0)

%     (3-0.5,2.0) to [short] (3+0.5,2.0)
%     (3-0.5,0.5) to [short] (3+0.5,.5)

%     (2.5,0.5) to [C, l=$C$] (2.5,2.0)
%     (3.5,0.5) to [L, l_=$L$] (3.5,2.0)
    
%     %(3.0,2.25) to [short,*-] (4.0,2.25) 
%     %to [short] (4.0,2.0) node[rground] {}

%     %(3.0,1.75) to [short,*-] (4.0,1.75) 
%     %to [short] (4.0,2.0) node[rground] {}
    
%     (3.0,2.25) to [short,-*] (4.0,2.25)
%     (3.0,0.25) to [short,-*] (4.0,0.25)
    
%     (4.0,2.4) to [open, v^>=$U$] (4.0,0.1)
  
%     ;

% \end{scope}

\end{tikzpicture}
\caption{\label{fig:resistive-cooling}
A moving trapped electron induces an image current in the trap electrodes, which creates a potential difference U and is dissipated via the impedance of an attached tank circuit which is held at 0.4\,K.}
% (b) The corresponding equivalent electronic circuit to simplify the analysis is shown.}
\end{figure}
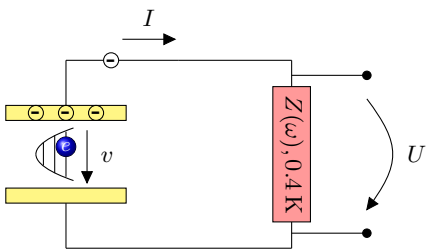

Trapping electrons and cooling them to low temperatures to minimize the average extent of their wavefunction, thus allowing for a non-destructive spin state readout, is essential for quantum computing with trapped electrons.
The electron motion can be damped by coupling the image current induced by the electron motion to a high-impedance cryogenic tank circuit, as shown in Fig.~\ref{fig:resistive-cooling}. This has been demonstrated in Penning trap experiments~\cite{Wineland1975}. 
The damping occurs because the current induced in the tank circuit by image charges from the moving electron is dissipated by the circuit's resistance, creating a voltage at the trap electrodes that opposes the electron motion until the induced current is of similar magnitude as the thermal (Johnson) current fluctuations in the circuit.
%Following the treatment in Ref.~\cite{Wineland1975}, we describe the electron as a series LC-circuit with inductance $L_e=\frac{m_e d^2_{\rm eff}}{e^2}$ and capacitance $C_e = \frac{1}{L_e \omega_{\rm t}^2}$ (see Fig.~\ref{fig:resistive-cooling}). 

The induced current is $I=ev/d_{\rm eff}$, where $v$ is the velocity of the electron with charge $e$ and $d_{\rm eff}$ describes the effective distance of the electrode structure \cite{Wineland1975,Brown1986a}. For the $y$-direction and the cooling circuit attached to the two center pickup electrodes on the top layer,  we find  $d_{\rm eff} = 138$\,\micrometer, a reduction by about a factor of two as compared to the ideal plate capacitor geometry. For the axial ($z$) direction with the cooling circuit attached to one of the top center pickup electrodes, we find $d_{\rm eff} = 254$\,\micrometer.

The cooling time constant can be derived by considering the electron as a circuit element~\cite{Wineland1975}
\begin{equation}\label{eq:cooling-time-constant}
\tau=\frac{m_e}{e^2}\frac{d_{\rm eff}^2}{{\rm Re}(Z)}\:.
\end{equation} Here, $Z = Q\sqrt{L/C}$ is the on-resonance impedance of the attached circuit. 
%We note that the cooling can be improved substantially by the high on-resonance impedance $Z=Q\sqrt{L/C}$ of a parallel $LC$-circuit with quality factor $Q$.
Assuming that the pick-up electrodes are attached to a tank circuit with a moderate quality factor of $Q = 1,000$, a capacitance of C=1\,pF, a resonant frequency of $2\pi\times 2$\,GHz, and an inductance of L=6\,nH results in 
% a characteristic impedance of $Z_0 = \sqrt{L/C} = 80$\,$\Omega$ and 
an on-resonance impedance of ${\rm Re}(Z) = 80$\,k$\Omega$.
Inserting this into \req{eq:cooling-time-constant}, we find a cooling time constant of 8\,\microsecond for the mode in the $y$-direction. For the mode in the $z$-direction, the cooling time constant is approximately 4\,\microsecond ($\omega_{\rm a}=2\pi\times300$\,MHz, $L=250\,$nH, $C=1\,$pF, $Q=1,000$, $Z=500\,$k$\Omega$). 

In equilibrium, the average energy of the cooled mode will be that of the temperature of the tank circuit, i.e.~$\braket{E} = \hbar \omega_{\rm t} (\bar{n} + 1/2) = \kB T_{\rm t}$. 
Assuming 2\,GHz for one of the transverse modes and $T_{\rm t} = 0.4$\,K, the average motional excitation will be $\bar{n} \approx 4$. 
%To achieve high-fidelity two-qubit gates, the low frequency 300\,MHz axial mode will need to be cooled. We plan to accomplish this by cooling first a transverse mode and then swapping the excitations of the transverse and axial mode. 

\label{sec:parametric-coupling}
Since the vertical transverse mode can be cooled following this protocol, the horizontal transverse mode and the axial mode can also be cooled by coupling them to the vertical transverse mode. 
Such parametric coupling is common practice for electrons and ions in Penning traps~\cite{Brown1986a} and has been demonstrated for ions in a Paul trap~\cite{Gorman2014}.
To achieve this, our electrode geometry allows us to generate a quadrupole field with projection overlap on a pair of modes that oscillates at the difference frequency between the two modes. This leads to an interaction that causes an energy exchange between the two motional modes. Assuming that we hold the high-frequency vertical transverse mode at 0.4~K, the low-frequency axial mode will be cooled to a temperature reduced by the frequency ratio,
$T_{\rm a} = T_{\rm t} \cdot \omega_{\rm a}/\omega_{\rm t} = 60\,{\rm{mK}}$. Scaling results from Ref.~\cite{Gorman2014}, we expect coupling times between the two motional modes of below 1~\microsecond.
% Thus, the cooling can be switched on and off at will. This is important as for the two-qubit gates and part of the state detection the axial motion should not be disturbed. 

Since resistive cooling does not affect the spin state of the qubit, the electrons can be cooled while keeping the qubits coherent, and no sympathetic coolant-particles are needed to be able to perform high fidelity operations on electron qubits. 
In fact, the cooling does not need to be turned off for most of the time: only during multi-qubit gates where a common mode of motion is used as a bus, as well as during the spin-motion conversion process for detection, must the cooling be switched off.
We also note that efficient cooling takes place only if the electron motion is on resonance with the resonant frequency of the tank circuit. 
Thus, we can adjust the cooling time constant by changing the trapping potential.

\subsection{Spin readout and initialization of electrons}
\label{sec:Spin readout of electrons}

The spin direction can be read-out by first coupling the spin to the electron motion and then measuring the phase of the motion, as discussed in Ref.~\cite{Peng2017}.
In order to couple the spin direction to the motion, one can apply a magnetic field gradient oscillating at the axial motional frequency. % near 300\,MHz. 
As the direction of the force due to the magnetic field gradient depends on the spin direction, this creates a state-dependent displacement acting on the initial thermal state.
% state of the form $a \ket{\uparrow}\ket{-\alpha}+b \ket{\downarrow}\ket{\alpha}$ where $\alpha$ describes the displacement of the resulting coherent state. 
%Once the displacement $\ket{\alpha}$ of the motion exceeds that of the random background excitation, due to the finite temperature of the electron, a parametric amplification scheme can be used to increase the displacements to easily measurable sizes by modulating the stiffness of the trap potential at twice the electron's motional frequency \cite{Burd2019,Burd2021}.
The measurement result is encoded in the phase of the motion, and thus the critical requirement is to maintain the coherence of the motion.
Therefore, it is important to create the displaced state faster than the motional dephasing time. Thus, one of the limiting factors is the strength of the magnetic field gradient.
%We find that for an axial temperature of 600\,mK as provided by coupling the axial motion to the transverse motion, and a magnetic field gradient of 91\,T/m,
We calculate that passing 1\,A of current through the hairpin wires in \rfig{fig:prelimanary-trap-design-2L} will create a gradient of
$120$\,T/m. Applying an oscillating current of this amplitude at the axial motional frequency of the electrons for $10\,\microsecond$ maps the spin direction to the motion with a fidelity of 99.7\,\%, assuming an initial axial mode temperature of 60\,mK, cooled via parametric coupling as described before.
To increase the amplitude of the image current further to a readily measurable size, electric forces can be used to parametrically amplify the coherent state $\ket{\alpha}$ by modulating the trapping voltages at twice the trap frequency~\cite{Burd2019,Burd2021}.

%Finally, we either detect the phase of the coherent signal with a tank circuit resonant with the axial motion, or we utilize the cooling tank circuit already in place by swapping oscillator states of the axial and radial motion with the parametric coupling already used for cooling. The latter interactions are all based on electric fields and hence, we expect that they can be carried out in less than 1~$\mathrm{\mu s}$. 
%Finally, image current detection takes only a few time constants. Thus, the time required for read-out will be dominated by the state-dependent force step estimated to take about 20~$\mathrm{\mu s}$.

Since it is critical to conserve phase coherence, it is necessary to consider the harmonicity of the axial trapping potential created by the DC electrodes in \rfig{fig:prelimanary-trap-design-2L}. 
Experimentally one can use the potentials applied to the twenty DC electrodes to minimize anharmonicities in the axial potential.  
% Assuming that the voltages on each electrode are provided by conventional $\pm$10\,V, 16-bit digital-to-analog converters, limiting the voltage resolution in this optimization step. t
Expanding the optimized trap potential into the Taylor series $V(z)=V(0)(1+c_2 z^2+c_4 z^4+c_6 z^6)$ and $c_2=1 $(\textmu{}m)$^{-2}$, we find coefficients $c_4=10^{-7}$(\micrometer)$^{-4}$ and $c_6=-2\times 10^{-9}$(\micrometer)$^{-6}$, while odd and higher even order terms are negligible, assuming a typical 16-bit DAC voltage uncertainty of 200\,\textmu{}V for each trapping voltage. From this we find the trap frequency uncertainty as 
$\Delta \omega_{\rm a}/\omega_{\rm a}\approx (3 A^2 c_4/4 + 15 A^4 c_6/16)/c_2$
, where $A$ is the amplitude of the motion.
To excite the motion above the Johnson noise at 0.4\,K, the amplitude must exceed $A>1.3$\,\micrometer. 
At this amplitude, the relative frequency shift due to the trap anharmonicity would be only $1.2\times10^{-7}$, corresponding to an absolute shift of the resonance frequency of 36\,Hz, which is negligible on the considered timescale of 10\,\textmu{}s.  

The current induced by the motion of the electron produces a voltage drop across the tank circuit that can be picked up with a Johnson-noise limited amplifier.
The phase of the electron motion is then encoded in the phase of the amplified voltage.
Electronic detection of electrons is standard in Penning traps~\cite{Wineland1973,Brown1986a,Cornell1990,Sturm2011}. In Paul traps, however, the strong trap drive of $\sim 14$\,V may saturate the amplifiers, requiring careful filtering (see \rapp{appendix: strong drive in motion detection}).

With state read-out in place, we can initialize electron spin by first measuring it and then flipping it conditioned on the measurement result. 
If the electron is found in the state we would like to initialize it in, nothing is done. If it is in the other state, we perform a $\pi$-pulse using a microwave field.

\label{sec:spin-initialization}

\subsection{Quantum gates with electrons}
%\label{sec:parametric-coupling}

Single-qubit gates can be performed by using microwave pulses near the Zeeman resonance, similarly to how error rates of $10^{-6}$ have been achieved for ions \cite{Harty2014}. 
For evaluating two-qubit gates, the M\o lmer-S\o rensen gate~\cite{Molmer1999,Roos2008-MS-gate} and its controlled phase gate  ($\sigma_z \otimes \sigma_z$) variant~\cite{Leibfried2003a} are considered. While typically performed using optical-frequency radiation, it can also be implemented using static or oscillating magnetic field gradients~\cite{Ospelkaus2011,Mintert2001a,Ospelkaus2008,Srinivas2018b,Webb2018,hahn_2019}. 
%Focusing on the oscillating magnetic field gradient version, 
The general idea is that a force oscillating nearly resonant with a mode frequency of the two-electron crystal excites the motion if the force on the individual electrons has the correct symmetry. For the phase gate variant, a gradient oscillating at $\omega_{\rm a}\pm \delta$ with $\delta \ll \omega_{\rm a}$ excites the center-of-mass motion of the electron crystal along the axial direction if the electrons are in the same spin-state, i.e.  $\ket{\uparrow \uparrow}$ or $\ket{\downarrow \downarrow}$. If the electrons are in the $\ket{\uparrow \downarrow}$ or $\ket{\downarrow \uparrow}$ state, the net force vanishes and the electron-COM (Center-Of-Mass) motion is not excited.
The corresponding Hamiltonian in the interaction picture is given by~\cite{Leibfried2003a,Roos2008-MS-gate}
\begin{equation}\label{eq:ms_hamiltonian}
% \hat{H} = \frac{\mu_{\rm B}}{2\sqrt{2}}\frac{d B}{dz}z_0(\hat{I}\otimes\hat{\sigma}_z+ \hat{\sigma}_z\otimes\hat{I})
\hat{H} = \hbar\Omega_{\rm R}(\hat{I}\otimes\hat{\sigma}_z+ \hat{\sigma}_z\otimes\hat{I})
(\hat{a}e^{-i\delta t}+\hat{a}^\dag e^{i\delta t} ),  
\end{equation}
where $\Omega_R$ is the two-qubit gate Rabi frequency, $\hat{a}$ is the motional mode lowering operator, and $\sigma_z$ is the Pauli  operator.
% , Fig.~\ref{fig:MS_gate} (b)
Because of the detuning $\delta$ of the force, the motion returns to its initial state for all four logical eigenstates after the gate time $t_{\rm gate}=2\pi/\delta$, enclosing an area in phase space. %The size of the area depends on the strength of the magnetic field gradient $\frac{d B}{dz}$ and the detuning $\delta$ of the magnetic field gradient drive from the secular frequency $\omega_{\rm a}$ (or from the motion dressed by the qubit, i.e. $\omega_{\rm a} + \omega_{\rm L}$, where $\omega_{\rm L}$ is the Larmor precession frequency of the electron in the static magnetic field).  
Different combinations of spin states acquire different geometrical phases, proportional to the area enclosed by their trajectories in phase-space. For our parameters, the anticipated gate time for a two-qubit phase gate is about $2\,\microsecond$.

%By choosing $\hbar \delta = \mu_{\rm B}\frac{d B}{dz}z_0$ the phase space trajectory closes at the gate time.
% defined by
% \begin{equation}\label{eq:gate-time}t_{\rm gate}
% = 2\hbar\pi/ \left(\mu_{\rm B}\frac{d B}{dz}z_0\right)\:.\end{equation}
%Under this condition, the motional mode is decoupled at the end of the gate and a spin entangled state is generated. 

For simplicity, we discussed the $\sigma_z \otimes \sigma_z$ gate, but $\sigma_x \otimes \sigma_x$ gates may offer certain advantages, as the drive is not near the motional frequency thereby suppressing unwanted excitation via residual electric fields. The $\sigma_z \otimes \sigma_z$ gate can also be
implemented with all fields far detuned from the motional
frequencies~\cite{Srinivas2021}. Such variants have been implemented using ions~\cite{Ospelkaus2011} with Bell state fidelities $F={\rm Tr} (\rho_{\rm exp} \rho_{\rm Bell})$ reaching 99.7\%~\cite{Harty2016}, limited by motional heating and trap frequency instabilities, and 99.9\%~\cite{Srinivas2021}, limited mainly by motional dephasing, where $\rho_{\rm exp}$, $\rho_{\rm Bell}$ are the density matrix operators of the experimentally prepared state and the ideal Bell-state intended to prepare respectively.

\section{Error sources of two-qubit gates}
\label{sec: Error souces}

\begin{figure*}[htb!]
\includegraphics[width=\textwidth]{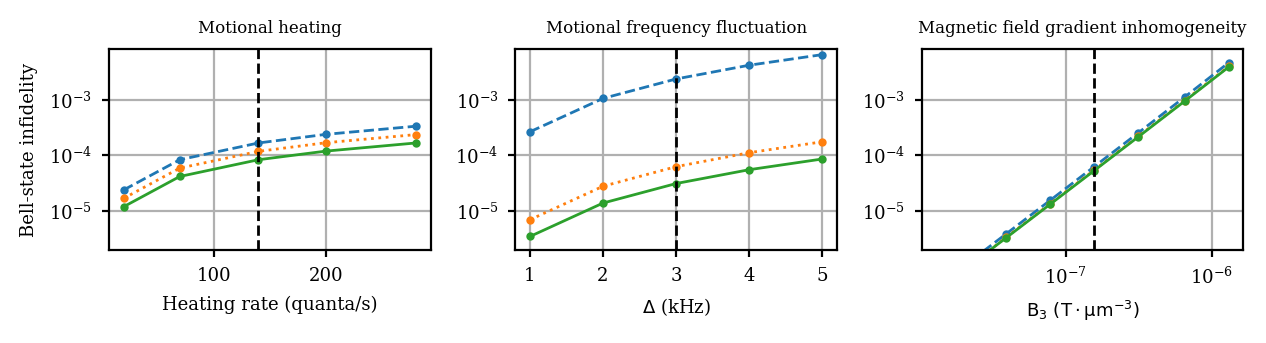}
    \centering
    \caption{Numerical simulation (QuTip \cite{johansson2012}) of three dominant error sources for an electron two-qubit gate. Blue dashed, orange dotted and green solid curves correspond to no modulation, Walsh 1 modulation and Walsh 3 modulation, respectively. For the magnetic field gradient inhomogeneity simulation, the three traces for different Walsh modulations overlap. To allow for an easy comparison, the $y$-axis is the same on all three plots. Black dashed lines indicate our anticipated magnitude of the different error terms. Based on these error estimates, we estimate that the anticipated contribution to the Bell-state infidelity can be limited to $10^{-4}$ for all error sources considered here.
    \label{fig:error-sources}
 %   \vspace*{-0.5cm}
    }
\end{figure*}

High-fidelity gate operations are crucial for achieving fault-tolerant quantum computing. In the following, we use Bell-state fidelity as a proxy for operation fidelities and analyze the most important sources of gate infidelities for electron qubits in our trap configuration.
Figure~\ref{fig:error-sources} shows an overview of the Bell-state infidelities ($1-F$) due to the most relevant decoherence sources (see \rapp{app:quantum_gate_simulation} for details).   
Many of these error sources change slowly compared to the gate time and can be taken as quasi-static. This opens up the possibility of using dynamical decoupling sequences (for example, Walsh modulation) as used in state-of-the-art  trapped-ion experiments ~\cite{Harty2016,Manovitz2017,Webb2018,Sutherland2019-laser-free,Sutherland2020,Srinivas2021}, with a modest overhead in gate duration.

%\begin{description}
%\item {\bf Motional heating } %(we need to use the COM mode because the other modes will likely have too much anharmonicities)
%{\bf Motional heating}
%is expected to be on the order of 14\,quanta/s (Eq.~\ref{eq:actual-heating}), corresponding to acquiring one motional excitation in 70\,ms. For anticipated gate times of 2~$\mathrm{\mu s}$, this 
% {\bf Motional heating:}
% \subsection*{Motional heating}

\subsection{Motional heating}
One of the dominant sources of error can be motional heating due to surface electric field noise. The motional heating rate is proportional to the electric-field spectral noise density $S(\omega_{\rm a})$ \cite{Brownnutt2015,Brown2021-materials-challenges}
\begin{equation}\label{eq:heating}
    \dot{n}=\frac{e^2}{4m_e \hbar \omega_{\rm a}}S(\omega_{\rm a})\:.
\end{equation}
For a 30-\micrometer ion-electrode distance, $S(\omega_{\rm a})$ near 1\,MHz is expected to be approximately $10^{-12}\,\mathrm{V}^2\mathrm{m}^2/\mathrm{Hz}$ at cryogenic temperatures \cite{Brownnutt2015}. This corresponds to a heating rate of $\dot{n}_{\rm Ca}=100$\,quanta/s at $\omega_{\rm a}^{\rm Ca} = 2\pi \times 1$\,MHz for Ca$^+$ ions. 
% 30 um
In order to extrapolate to GHz frequencies, we assume that the spectral noise density follows a power law $S(\omega_{\rm a})\propto 1/\omega_a^\gamma$. For ion traps, values of $\gamma$ ranging from 1 to 1.5 have been measured~\cite{Hite2012,Sedlacek2018-multi-mechanisms,Noel2019-TAF,Brown2021-materials-challenges}. Taking into account the light mass of electrons and an axial frequency of $\omega_{\rm a} = 2\pi \times 300$\,MHz, the expected motional heating rate for electrons is 
%$\dot{n}_{\rm e}$ is
%\begin{equation}     
%\dot{n}_{\rm e}=\frac{m_{\rm e}}{m_{\rm Ca}}
%\frac{\omega_{\rm Ca}^2}{\omega_{\rm e}^2} \approx 100\:~{\rm quanta/s}\:, \end{equation}
\begin{equation}\label{eq:actual-heating}     
\dot{n}_{\rm e}
=
\frac{m_{\rm Ca}}{m_{\rm e}}
\left(\frac{\omega_{\rm a}^{\rm Ca}}{\omega_{\rm a}}\right)^{1+\gamma}\dot{n}_{\rm Ca}\:.\end{equation}
%\approx 100\:\,{\rm quanta/s}\:, 
For niobium traps at cryogenic temperatures, an exponent $\gamma=1.3$ has been found~\cite{Sedlacek2018-multi-mechanisms}, leading to an estimate for the heating rate of 14\,quanta/s, which corresponds to a limit of the motional coherence time of 70\,ms. 
Because of the uncertainties of extrapolation over two orders of magnitude in frequency, we conservatively assume a heating rate of 140\,quanta/s, which yields a simulated Bell-state infidelity of about $8\times 10^{-5}$ with Walsh 3 modulation, as shown in the left panel of Fig.~\ref{fig:error-sources}. 
For the transverse modes near 2\,GHz, motional heating should be reduced by approximately two orders of magnitude to order 1\,quanta/s.  
% In addition, milling the surface with noble-gas ions may reduce electric field noise at cryogenic temperature further~\cite{Sedlacek2018-multi-mechanisms}.

%Further we note that it is reasonable to expect that the electric field noise from the trap electrodes is an equilibrium effect and one would expect that as such it would tend to drive the electron temperatures towards the temperature of the trap electrodes.
%The heating# in the axial direction may  impact gate fidelities. We will analyse this in detail later, but state here that for the  anticipated gate times of 100\,ns, motional heating is expected to increase the error rate by less than $10^{-5}$.
%Furthermore, reduction of the impact of the heating rate on the gate infidelities can be expected when using advanced dynamical decoupling sequences as demonstrated for ions~\cite{Webb2018,Manovitz2017}

% Further resilience can be obtained by using dynamical decoupling schemes as already demonstrated for ions~\cite{Manovitz2017,Webb2018}. 

%\item {\bf Trap frequency instabilities: }
% {\bf Trap frequency instabilities:} 

\subsection{Trap frequency instabilities}

Instabilities of the trap frequency are caused by voltage noise, finite temperature of the electrons in combination with anharmonic potentials, as well as surface electric field noise.
Trap frequency instabilities lead to a fluctuating detuning of the drive frequency from the motional sidebands at $\omega_{\rm a}$, and thereby to errors.

Formally, we distinguish between fast noise (dephasing) and slow noise (trap frequency fluctuations) as compared to a gate duration. Slow fluctuations can be mitigated effectively using modified phase space trajectories \cite{hayes2012coherent,Sutherland2020,Haddadfarshi2016,Webb2018,Shapira2018}. Fast noise, on the other hand, cannot be easily mitigated other than by speeding up the quantum gates or by performing more loops in phase space. 

Trap frequency fluctuations due to noise of the voltages applied to the electrodes are expected to be slow if low-pass filtered. 
Typical voltage sources provide fractional voltage stabilities on the order of $10^{-5}$, corresponding to a motional frequency fluctuation $\Delta = 2\pi \times 3\ \mathrm{kHz}$.
Using Walsh modulation one can suppress errors from slow frequency noise\cite{hayes2012coherent}. Assuming the above numbers, we estimate a residual Bell-state error of $3\times10^{-5}$ with Walsh 3 modulation, as shown in the center panel of Fig.~\ref{fig:error-sources}.

Finite electron temperatures lead to fluctuations in their oscillation periods due to the trap potential anharmonicities. Since motional heating is negligible over gate durations, this noise is also slow and can be taken care of by using Walsh modulations.

The quadrupole component of the surface electric field noise introduces noise to the quadrupole trapping potential, thereby causing fluctuations of the trap frequency. This noise component is related to the surface electric field noise as $S_{\rm Q}(\omega_{\rm a}) = \frac{15}{4 d^2} S(\omega_{\rm a})$, where $d$ is the electron-electrode distance, and $S(\omega_{\rm a})$ exhibits a $1/f^\alpha$~\cite{Talukdar2016} characteristics with frequency scaling exponent $\alpha \gtrsim 1$, consisting of both fast and slow noise components, where $f$ is the frequency of the noise.
Ref.~\cite{Talukdar2016} estimates that at 
$d\approx 25$\,\micrometer the integrated impact of quadrupole noise on the Bell-state fidelities is similar to that of motional heating.
%, i.e.\,$T_1 \approx T_\phi$.
However, since the bulk electric field noise has a $1/f$ scaling, it can also be reduced by Walsh modulations.

In order to estimate the size of the fast noise, we draw on the insights of a recent ion trap experiment where for a trapping height of 30\,\micrometer its impact has been estimated to contribute $6\times10^{-4}$ to the Bell-state infidelity~\cite{Burd2021,Srinivas2021}. 
%In these experiments gate times were on the order of 1\,ms, while we anticipate gate times of order 2~$\mathrm{\mu s}$. Further, the noise is expected to scale with the frequency as $1/f$. 
The gate time in these experiments was several hundred microseconds, more than two orders of magnitude slower than what we expect for electrons. Thus, we expect that a significant amount of the fast noise observed in the ion trap experiments is actually slow in the context of electron quantum computing and can be removed with Walsh modulations. 
Assuming a $1/f$ characteristic, the remainder of the noise leads to a dephasing rate of $\Gamma = 2\pi \times 1.8\times 10^{-3} \ \mathrm{Hz}$, which corresponds to a Bell-state infidelity of $5\times10^{-8}$.

%In order to estimate effect of  fast noise in general, we draw upon work done by the NIST group estimating the contribution of dephasing to their Bell-state fidelity from their squeezing experiments~\cite{Burd2021} to $6\times10^{-4}$ (see Sec. 4.1 in the supplement of Ref.~\cite{Srinivas2021}). In these experiments gate times were on the order of 1\,ms, while we anticipate gates of order 2~$\mathrm{\mu s}$. Further, the noise is expected to scale with the frequency as $1/f$. Thus, we expect that most of noise present in Ref.~\cite{Srinivas2021} is slow as compared to the entangling gate time and can be efficiently mitigated using dynamical decoupling. Assuming a $1/f$ characteristics, the remainder of the noise should contribute to a gate infidelity of  $2\times10^{-6}$.  

\subsection {Inhomogeneity of the magnetic field gradient}

%One difference between trapped electrons and ions is that the motional amplitude of the electrons at finite temperatures tends to be on the order of 1~$\mathrm{\mu m}$, much larger than for trapped ions. 

The motional amplitude of trapped electrons at 0.4~K is a few hundreds of nanometers, which is large compared to $\sim$ 10\,nm for ions. 
As a result, error sources usually not considered for trapped ions may become relevant. In particular, the electron experiences a variation of the force over the extent of its wavefunction, due to any inhomogeneity of the magnetic field gradient. 
Thus, for electron qubits, the homogeneity of the applied microwave radiation in combination with the finite electron temperature is relevant.
%This leads to noisy coupling to the microwave drive since the large motional amplitude shifts lead to thermally randomized forces on the electron due to variations in the magnetic field gradient. 

%This relatively large motional amplitude leads to noisy coupling to the drive field if the magnetic field gradient is not homogeneous. The reasoning is that thermally randomized amplitude of the electron motion of order 1~$\mathrm{\mu m}$ randomizes the force on the electron due to variations of the magnetic field gradient over this distance. Thus for electron qubits, the homogeneity of the applied microwave radiation in combination with the finite electron temperature is of concern. 

%The size of this effect depends critically on the temperature of the electrons and the relative strength of the magnetic field gradient inhomogeneity. 

To estimate the lowest order effect of inhomogeneity, we do a Taylor expansion of the magnetic field, as discussed in \rapp{app:quantum_gate_simulation}. Due to symmetry, the contribution of the second order term (corresponding to the first order term in magnetic field gradient $\nabla B$) cancels over the extent of the electron motion. The third order term (corresponding to the second order term in $\nabla B$), on the other hand, introduces a loss of fidelity due to a non-cancelling force. 
Using the hairpin design shown in \rfig{fig:prelimanary-trap-design-2L}, we find the third order expansion coefficient of the magnetic field to be $B_3=1.5\times 10^{-7}$\,T/\textmu{}$\rm{m}^3$, which causes an infidelity of $5\times10^{-5}$ with Walsh 3 modulation (see the right panel of Fig.~\ref{fig:error-sources}).
%We anticipate that by optimizing the geometry of the hairpin design and/or adding additional wires, the homogeneity can be substantially improved. 
Finally, we note that this effect can be reduced by reducing the temperature of the electrons as discussed in \rapp{app:thermal_load}.

%we find for the third order expansion coefficient of the magnetic field $B^{(3)}=5\times 10^{-7}$\,T/${\mathrm{\mu m}}^3$ for the leading order $z^3$ in the magnetic field itself (

%corresponding to $2\times 10^{-7}$~T/${\mathrm{\mu m}}^2$ for the quadratic order in $\nabla B$), which causes a $6\times10^{-5}$ fidelity loss. We anticipate that by optimizing the geometry of the hairpin design and/or adding additional wires, the homogeneity can be substantially improved. Finally, this effect can also be reduced by reducing the temperature of the electrons. 

\subsection{Anharmonicity of the trapping potentials}

%The anharmonicity of the Coloumb interaction between the electrons leads to cross-mode coupling \cite{Roos2008c,Nie2009}. 
%The anharmonicity of the Coulomb-interaction between the electrons leads to frequency fluctuations of all but the center-of-mass modes \cite{Roos2008c,Nie2009}. 
%To counter this, one typically cools all modes near the motional ground state. To avoid the complications of ground state cooling, one can use the center-of-mass mode, which is immune to this effect. 
% Further, this noise is slow and would be mitigated by our dynamical decoupling methods.

The effect of the temperature on the COM mode frequency is given solely by the anharmonicity of the trapping potential itself. 
As discussed in \rsec{sec:Spin readout of electrons}, with the optimized trap potential up to the fourth order $V(z)=V(0)(1+c_2 z^2+c_4 z^4)$, we find $c_4/c_2=10^{-7}$(\textmu{}$\rm{m})^{-2}$, which corresponds to only very small contributions to the Bell-state infidelity and is on the order of $10^{-7}$ or below for Walsh modulations.

\subsection {Qubit decoherence}

The decoherence of the electron spin is expected to be dominated by magnetic field noise. 
Using magnetic shielding, coherence times of 300\,ms (2\,s with a spin-echo) have been observed for Zeeman qubits in ions~\cite{Ruster2016}. 
The greatly reduced optical overhead requirements for electron qubits as compared to the experiment in Ref.~\cite{Ruster2016} will allow for more efficient magnetic shielding, leading to expected spin-coherence times in excess of 1\,s without spin-echo.
This corresponds to only small contributions to Bell-state infidelities on the order of $3\times10^{-6}$.

In laser-free trapped-ion entangling gates, qubit frequency shift error can be suppressed effectively in a dynamically decoupled M\o lmer-S\o rensen gate~\cite{Harty2016} and $\sigma_z \otimes \sigma_z$ gate with intrinsic dynamical decoupling~\cite{Srinivas2021}. We expect a similar performance for electron qubits.

\begin{table*}[ht]
\footnotesize
\centering
\captionsetup{justification=centering}
\def\arraystretch{1.3}
\begin{tabular}{|c|c|c|c|c|c|c|}
\hline
      & \begin{tabular}[c]{@{}c@{}}Motional\\heating\end{tabular} &
      \begin{tabular}[c]{@{}c@{}}Trap frequency \\ fluctuation\end{tabular} &\begin{tabular}[c]{@{}c@{}}Motional\\dephasing\end{tabular}  &  \begin{tabular}[c]{@{}c@{}}Magnetic field \\ gradient inhomogeneity\end{tabular} & \begin{tabular}[c]{@{}c@{}}Trapping potential \\ anharmonicity\end{tabular}& \begin{tabular}[c]{@{}c@{}}Qubit\\decoherence\end{tabular} \\ \hline Magnitude  & 140\,quanta/s & 3\,kHz&$1.8\times 10^{-3}$\,Hz &$ 1.5\times 10^{-7}$\,T/\textmu $\rm{m}^3$ & $10^{-7}$\,\textmu{}$\rm{m}^{-2}$  & 1\,s\\ 
 Infidelity & $8\times 10^{-5}$ & $3\times 10^{-5}$ & $5\times 10^{-8}$   & $5\times 10^{-5}$ & $2\times 10^{-7}$ & $3\times 10^{-6}$\\[0.5ex] \hline
\end{tabular}
\caption{Estimates for the contribution of various error sources with Walsh 3 modulation. }
\label{table:error source}
\end{table*}

\subsection{Summary of decoherence sources}

In summary, we find that the largest source of infidelities will likely be magnetic field gradient inhomogeneities and motional heating, both expected to lead to errors smaller than $10^{-4}$ as shown in \rtab{table:error source}.  
While these infidelities meet some commonly assumed error correction thresholds, they can be improved further. 
For instance, surface treatments may reduce motional heating~\cite{Sedlacek2018-multi-mechanisms} and thus suppress one of the largest error sources. 
Another path towards reducing motional heating would be to increase the electron-electrode distance. However, this will compromise achievable magnetic field gradients and motional frequencies. 
%Yet another way to reduce errors is to cool the electrons transverse mode substantially below 0.4\,K with a dilution refrigerator, as discussed in \rapp{app:thermal_load}. This would lower the axial temperature well below 60\,mK.

%One may be concerned here by the heat-load considerations, however, only the ohmic resistance of the cooling circuit determines the temperature that the electrons would equilibrate at. \label{sec:thermal-decoupling} By thermally decoupling this circuit, we may cool it below 40\,mK with a dilution refrigerator stage as done in Penning trap experiments to determine the electron's g-factor. This would reduce the error sources related to motional heating, motional dephasing, as well as errors  due to magnetic field inhomogeneities by more than one order of magnitude. 
%Further, since electrons do not posses optical transitions, many fundamental error sources for trapped ions are not a concern for electron qubits such as spontaneous scattering of photons.

%\subsection{Error correction codes}
%required gate fidelity for fault-tolerant computing?

%%%%%%%%%%%%%%%%%%%%%%%%%%%%%%%%%%%%%%

\section{Modularization and Qubit Addressing}

A modular approach is preferred for constructing a large-scale quantum computer, and the most promising strategy for scaling trapped electron systems is adapting the quantum charge-coupled device (QCCD) architecture from trapped ions~\cite{Kielpinski2002CCD,Wineland1998,Lekitsch2017}. Since the trap frequencies for electrons we consider are approximately two orders of magnitude larger than typical values for trapped ions, we expect that electron transport and operations such as splitting and merging of the electron crystals can be carried out substantially faster than on equivalent trapped-ions QCCD architecture. 

Combining two-qubit gates with shuttling operations, we imagine entangling two electrons at specific sites in the QCCD device. Shuttling and splitting operations on electron crystals can be used to isolate a selected pair of electrons in a region where a localized magnetic field gradient can be applied to perform gates. These processing sites, with current-carrying wires providing the necessary magnetic-field gradient, would be on the order of 200\,\micrometer long to allow for high-fidelity single-qubit addressing. In addition, the trap frequency would serve as an important discriminating element. Thus, we expect that crosstalk for multi-qubit  gates can be even lower than that for single-qubit gates.
The electrons could be more densely packed in dedicated storage regions and can be shuttled between these two types of regions in a conveyor-belt fashion, and junctions connect different processing and storage units with each other on a 2D-grid to create large entangled states. 

To build a modular electron quantum computer, coupling of remote electron qubits could be engineered with a high-impedance co-planar waveguide that distributes image currents between distant sites.
In Ref.~\cite{Daniilidis2013electron}, the authors estimate such a coupling between two electrons to be on the order of 100\,kHz, which is a significant improvement over current implementations of remote entanglement in trapped ion systems~\cite{Hucul2014ModularPhonons,Stephenson2020-entanglement}.

%An electron QCCD platform

%\section{Scalability}
% \section{Outlook: Scaling up electron Paul traps}
\label{sec:scalability}

\section{Conclusion and outlook}
\label{sec:conclusion}

In summary, we have conducted a feasibility study of a trapped-electron quantum information processing platform and have discussed the experimental steps and potential challenges towards building such a device, including schemes for trapping, cooling, electronic detection, spin readout and quantum gates of electrons.

%After trapping electrons in a millimeter-sized Paul trap in room-temperature, the next experimental milestone is cooling electrons in a miniaturized Paul trap in cryogenic temperature and detecting their motion by measuring the induced image currents via an attached tank circuit. 
%The next capstone of this technology will be to detect electrons' spin direction and to perform single-qubit and two-qubit operations. 
Numerical simulations of quantum gates with electron qubits show that the largest sources of infidelities will likely be motional heating and magnetic field gradient inhomogeneities, both expected to lead to Bell-state preparation infidelities below  $10^{-4}$. 
This may meet the requirements for fault-tolerant quantum computing with reasonable overheads.
This system can be scaled up and modularized by adapting the QCCD architecture similarly to trapped ion system, allowing for single-qubit addressing and isolation of multi-qubit operations from bystander electrons. 
Therefore, the trapped-electron technology proposed here meets the DiVincenzo criteria \cite{divincenzo2000physical} for the physical realization of a quantum computer. 
%Possible challenges?

%Outlook part?

In addition to applications to quantum information processing, the development of techniques to trap cold electrons in Paul traps may also directly impact other disciplines, such as plasma physics through the study of small cold plasma \cite{twedt2012} and electron-positron interactions \cite{Surko2015}, 
% by using cold electrons for high-resolution imaging \cite{PhysRevA.80.040902}, 
and to serve as sensitive detectors of charges including millicharged dark matter \cite{carney2021,budker2021millicharged}. 
% The techniques developed here could also aid in the search for physics beyond the standard model through the preparation of cold positrons in a Paul trap for use in creating large amounts of anti-hydrogen \cite{Leefer2016}.

%Finally, our study concludes that it is feasible to adapt trapped-electron technology to quantum information processing.
%, and experimental investigations towards building a trapped electron quantum computer are encouraged.

%In conclusion, we laid out a prototype for a trapped-electron quantum computer that satisfies the DiVincenzo Criteria \cite{divincenzo2000physical} and operates at a fault-tolerant threshold.

\begin{appendix}

\section{Thermal loads on a cryogenic trap}
\label{app:thermal_load}

The model system we consider in Fig.~\ref{fig:prelimanary-trap-design-2L} is assumed to reside in a cryogenic environment at 4\,K with the resonant cooling circuit, as discussed further below, to be held at 0.4\,K. 

The most dominant heat source for the cryogenic electron trap is Joule heating from the current-carrying wires and AC-electrodes.
%External black-body radiation can be effectively shielded and almost completely suppressed because the operation of the electron trap requires only very minimal optical access.
To determine the heat load on the 4\,K stage, we assume that the current-carrying wires going from a 30\,K stage to the 4\,K stage are made of niobium with a length of 10\,cm, a diameter of 1\,mm, electrical resistivity of 1.5\,$\rm{n\Omega m}$ at 30\,K and thermal conductivity of 10\,$\rm{W/(m\cdot K)}$ at 4\,K. From this we estimate the heat load on the AC-electrodes to be $\approx 4.5$\,mW per 1\,A of current at 300\,MHz with a duty cycle of $\approx\,10$\,\% and $\approx 250$\,mW per 1\,A of current at 10\,GHz. More important will be the heat load from Joule dissipation in the AC-electrodes themselves, which is  directly transferred to the substrate. Assuming AC gradient electrodes with a cross section of $10\times 0.5\,\micrometer^2$ and a total length of 5\,mm operating with a duty cycle of $\approx\,10$\% for spin-readout and gates, the estimated heat load is $\approx 100$\,mW per 1\,A of current for copper with a specific resistance of $\approx 1\  \textrm{n}\Omega\textrm{m}$ and skin depth of $\approx$ 0.9\,\micrometer at 4\,K, 300\,MHz. Regarding the AC-trapping electrodes, assuming a drive voltage of 14\,V at 10\,GHz, and a capacitance of 1\,pF, this will lead to a current of 1\,A. Assuming further that the current flows through the entire AC copper electrode of length 5\,mm, cross section $50\times 0.5 \,\micrometer^2$, and a skin depth of $\approx$ 0.16\,\micrometer, we arrive at a heat load $I^2R=600$\,mW per AC electrode pair, i.e. 1.2\,W for all four. 
%For Joule heating, we analyze the heating effects of the AC-current carrying wires within the trap. Though this current will only be pulsed with a duty cycle of $\approx\,10-20$\% in typical experimental sequences, we expect that it will be the most dominant heat source in the system. Since the hairpin will carry 1~A of current, we analyze the heat produced by this which is estimated to be $\approx$350$\mu$W.
This heat load can be mitigated by providing a sufficiently large heat sink to a typical cryo pulse-tube cooler with 1.8\,W of cooling power at 4\,K by carefully choosing the trap substrate with high heat conductivity (e.g.~sapphire) and trap mounting materials.

In order to reach 0.4\,K in the motion of the trapped electron, the resonant circuit must be cryogenically cooled to the same temperature.
This can be done by thermally anchoring the resonant circuit at 0.4\,K and connecting it to the pickup electrodes at 4\,K with a Nb (Niobium) wire. The wire needs to be a good electrical conductor and a good thermal insulator to ensure that the heat load of the trap on the resonant circuit is lower than the maximum cooling power of the 0.4\,K stage, $\approx$ 1.4\,mW. Coupling the resonant circuit with a Nb wire of 1\,cm length and 100\,\micrometer diameter, the heat load on the resonant circuit is $\approx\,150$\,\textmu{}W, which is within the cooling capacity of a commercial dilution refrigerator with a maximum cooling power of $\approx$ 1.4\,mW.

\section{Strong drive in motion detection}
\label{appendix: strong drive in motion detection}

The amplifiers needed to detect the electron signal are prone to saturation due to the strong AC-drive of the electron trap.
%, thus we need an effective filtering of the 10 GHz drive.
To estimate the degree of filtering necessary to successfully detect an electron in the presence of the 10 GHz drive, we model the electron as a Johnson-noise source with a bandwidth of 1\,MHz at 4\,K corresponding to a noise power of -142\,dBm, where the bandwidth is determined by the coupling of the electron to the tank circuit.
% $\gamma=1/\tau$. 
% This signal corresponds to a voltage of order 100\,nV at the resonators impedance of 100\,$\Omega$. 
Assuming a dynamic range of the amplifier of 70\,dB, the pick up at the drive frequency at $\sim10$\,GHz must be smaller than -70\,dBm. We estimate that the power required to drive the trap will be of order 1\,W (30\,dBm) requiring an effective filtering of 100\,dB. 

First, we note that the signal is not applied to the detection circuit. Secondly, the resonant circuit picking up the signal from the electron itself serves as a filter: the trap drive is many GHz detuned from the resonant circuits narrow resonance $\Delta\omega$ at $\omega=2\pi\times300$\,MHz. Further, the amplifier itself will be optimized to amplify the 300\,MHz signal rather than the high frequency drive. 
%Further, we plan to pick up the image current from electrodes in the top substrate in
%Fig.~\ref{fig:prelimanary-trap-design-2L} rather than in the bottom substrate thereby rejecting most of the voltage in the first place.
Finally, we can filter the signal of $\sim$300\,MHz, either via low pass filter, or if that is still not sufficient with another resonant circuit.

\section{Quantum gate simulation}
\label{app:quantum_gate_simulation}
The Bell-state fidelity is defined as
\begin{equation}F={\rm Tr} (\rho_{\rm exp} \ket{\Psi_{\rm Bell}} \bra{\Psi_{\rm Bell}})\end{equation}
% \begin{equation}F={\rm Tr} (\rho_{\rm exp} \rho_{\rm Bell})\end{equation}
% \begin{equation}\mathcal{F}=\bra{\Psi_{\rm Bell}}\rho_{\rm exp} \ket{\Psi_{\rm Bell}}\end{equation}
% where $\ket{\Psi_{\rm Bell}}$ is the targeted Bell-state and $\rho_{\rm exp}={\rm Tr}_{\rm m}(\rho_{\rm T})$ is the reduced density matrix of total system $\rho_{\rm T}$ tracing out the motional degree of freedom. The Bell-state infidelity is calculated as $1-\mathcal{F}$. $\rho_{\rm exp}={\rm Tr}_{\rm m}(\rho_{\rm T})$
% where $\rho_{\rm Bell}$ is the density matrix of the targeted Bell-state and $\rho_{\rm exp}={\rm Tr}_{\rm m}(\rho_{\rm T})$ is the reduced density matrix of total system $\rho_{\rm T}$ tracing out the motional degree of freedom. The Bell-state infidelity is calculated as $1-\mathcal{F}$.
where $\ket{\Psi_{\rm Bell}}$ is the state vector of the ideal targeted Bell-state and $\rho_{\rm exp}$ is the reduced density matrix of total system tracing out the motional degree of freedom. The Bell-state infidelity is calculated as $1-F$.

Gate dynamics are simulated with the Lindblad master equation using QuTip~\cite{johansson2012}:
\begin{equation}\dot {\rho}(t)
=
-\frac{i}{\hbar}[\hat H,\rho]+\sum_n \left[\hat L_n\rho \hat L_n^{\dagger}-\frac{1}{2}\rho \hat L_n^{\dagger}\hat L_n-\frac{1}{2} \hat L_n^{\dagger}\hat L_n \rho\right]
\end{equation}
where $\hat L_n$ is a Lindblad operator, $\hat H=\hat H_g +\hat H_e$ is the total Hamiltonian consisting of the gate dynamics $\hat H_g$ and possible error sources described by $\hat H_e$.

The gate Hamiltonian is described as:
\begin{equation}\label{eq:gate_hamiltonian}
\hat{H_g} = \hbar\Omega_R(\hat{I}\otimes\hat{\sigma}_z+ \hat{\sigma}_z\otimes\hat{I})
(\hat{a}e^{-i\delta t}+\hat{a}^\dag e^{i\delta t} )
\end{equation}

where $\Omega_R$ is the two-qubit gate Rabi frequency.

{\bf Motional heating:}
Motional heating is modeled with the Lindblad operators $L_1=\sqrt{\gamma }\hat a$ and $L_2 = \sqrt{\gamma }\hat a^{\dagger}$, where $\gamma = \dot{n}_e$ is the motional heating rate, $\hat a^\dagger$ and $\hat a$ are the creation and annihilation operator of the motional mode, respectively.

{\bf Motional frequency fluctuation:}
Motional frequency fluctuations are modeled with the Hamiltonian
\begin{equation}
\hat{H_e} =\hbar \Delta 
\hat a^\dagger \hat a
\end{equation} where $\Delta$ is the motional frequency detuning from $\omega_a$.

{\bf Motional dephasing:}
Motional dephasing is modeled with the Lindblad operator
$L_3=\sqrt{\Gamma} 
\hat a^{\dagger}
\hat a$, where $\Gamma$ is the dephasing rate of the electron motion.

{\bf Inhomogeneity of the magnetic field gradient:}
Taylor expanding the magnetic field at the center-of-mass of the two-electron crystal $B=B_0+B_1 z+B_2 z^2+B_3 z^3$, where $B_j$ denotes the $j^{th}$ order expansion coefficient of the magnetic field. Keeping only the slowly oscillating terms rotating $\propto \delta$, the magnetic field gradient inhomogeneity error is modeled with the Hamiltonian~\cite{sutherland2021one}:
% Taylor expanding the magnetic field gradient at the center-of-mass of the two-electron crystal $\nabla B=B^{(1)}+B^{(2)}\delta z+\frac{1}{2} B^{(3)} \delta z^2$, where $B^{(j)}$ denotes the $j^{th}$ order derivative of the magnetic field. Keeping only the slowly oscillating terms, the magnetic field gradient inhomogeneity error is modeled with the Hamiltonian:
\begin{equation}
\label{eq:10}
\begin{split}
\hat{H_e}=3\hbar \Omega_{\rm in} (\hat I\!\otimes\! \hat \sigma_z\!+\!\hat \sigma_z\! \otimes\!\hat I)
(\hat{a}\hat{a}^{\dagger}\hat{a}e^{-i\delta t}\!\!+\!\hat{a}^{\dagger}\hat{a}\hat{a}^{\dagger} e^{i\delta t})
\end{split}
\end{equation}
% where $\Omega_{\rm {in}} = \Omega_R\cdot \frac{j_0^2}{2}\frac{B^{(3)}}{B^{(1)}}$, and $j_0$ is the ground state extension of motional mode $j$.
where $\Omega_{\rm {in}} = \Omega_R\cdot 3z_0^2 B_3/B_1$, and $z_0$ is the ground state extension of axial motional mode.

{\bf Anharmonicity of the trapping potentials:}
Expanding the trapping potential to only the fourth order $V(z)=V(0)(1+c_2 z^2+c_4 z^4)$, the trap potential anharmonicity can be modeled with the Hamiltonian
\begin{equation}
\hat{H_e} =V_4\hat z^4=V(0)c_4 z_0^4(\hat a_z+\hat a^\dagger_z)^4
\end{equation}

{\bf Qubit decoherence:}
Qubit decoherence is modeled with the Lindblad operator $\hat L_4=\sqrt{\frac{1}{2\tau_{\rm spin}}}\hat{\sigma}_z$, where $\tau_{\rm spin}$ is the spin coherence time.

\section{Electron trajectory stability}
\label{appendix: trap stability}
\begin{figure}[htb!]
\includegraphics[width=0.5\textwidth]{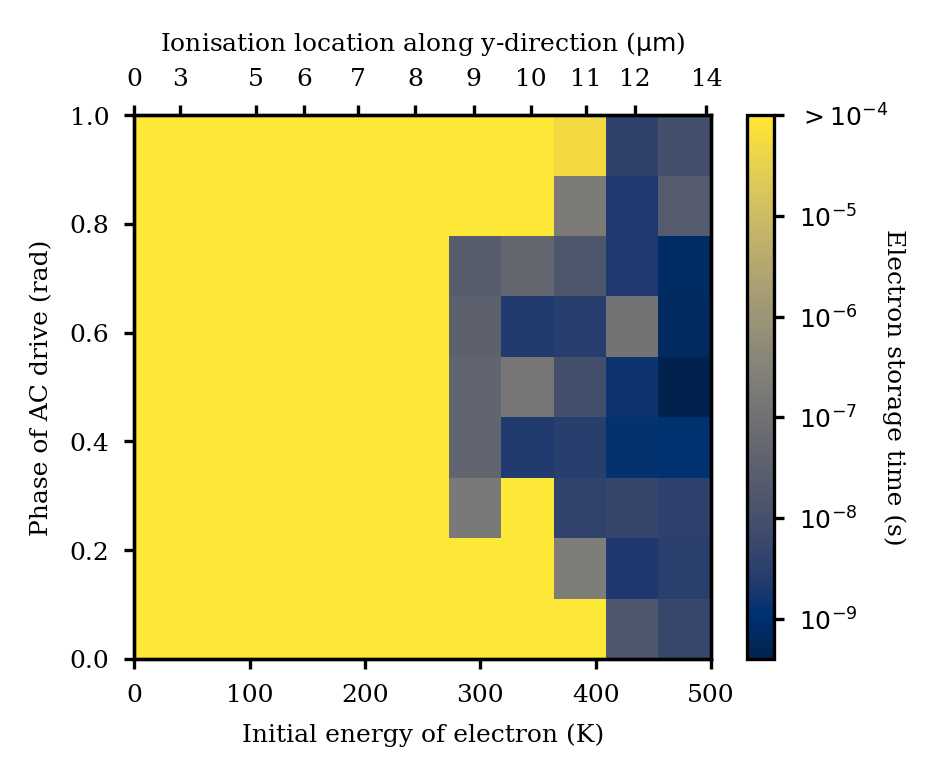}
    \centering
    \caption{Electron trajectory simulation of the two-dimensional secular motion. Electron trajectories are stable irrespective of the drive phase when the electron separates from the atom for initial energies less than 250 K. 
    \label{fig:electron trajectory simulation}
 %   \vspace*{-0.5cm}
    }
\end{figure}

The electron trajectory stability is analyzed by numerically integrating the two-dimensional electron motion along radial directions. Since the surplus ionization energy is small compared to the potential energy in the trapping potential, we assume the initial kinetic energy of electrons to be zero. The simulation variables are the phase $\phi$ of the AC drive at the ionization time ($t=0$) assuming a time dependence of $\rm {cos}(\omega_{\rm {ac}} t+\phi)$ and the initial energy of electrons in the AC trapping potential which is determined by the ionization
distance from the trap center. Fig.~\ref{fig:electron trajectory simulation} shows the simulation result, which is a map of the electron storage time simulated up to 100~\microsecond. From the simulation, electron
trajectories are not lost over the duration of the simulation for initial energies less than $k_{\rm B} \times 250$~K which corresponds to an ionization distance of about 8~\micrometer from the trap center. For higher energies, the stability of the electron trajectory depend strongly on the phase of the AC drive, and electron trajectories are the most stable when the phase of the AC electric field is zero when the electron is separated from its parent atom. 

% The numerical simulation is carried out assuming that electrons start at certain location with zero initial velocity and a certain initial phase of the AC drive. One may be concerned that as electron oscillates in the trapping potential, when electron reaches the turning point of its motion and the velocity becomes zero, it will turn into different initial condition settings with AC drive phases where the trajectories are less stable.  
% As shown in Fig.~\ref{fig:relative phase}, for initial conditions with stable trajectories, the relative phase between the electron secular motion and the AC drive is always multiples of $\pi$ at the turning point of the secular motion. Therefore, without perturbations, electron trajectory stability is maintained.

% \begin{figure}[htb!]
% \includegraphics[width=0.45\textwidth]{FigD2.png}
%     \centering
%     \caption{Relative phase between electron secular motion and the AC drive at turning point for different initially stable trajectories. The relative phase between electron secular motion and the AC drive is always zero when electron secular motion reaches the turning point. 
%     \label{fig:relative phase}
%  %   \vspace*{-0.5cm}
%     }
% \end{figure}

\end{appendix}

\begin{acknowledgments}
We acknowledge support from AFOSR through grant FA9550-20-1-0162, the NSF QLCI program through grant number OMA-2016245. Work by A.A. and H.H. was sponsored by the U.S. Department of Energy, Office of Science, Office of Basic Energy Sciences under Award Number DE{-}SC0019376. Q.Y., K.M.B. and R.T.S. acknowledge funding from the Lawrence Livermore National Laboratory (LLNL) Laboratory Directed Research and Development (LDRD) program under Grant No. 21-FS-008. 
Work done by K.M.B. was performed under the auspices of the U. S. Department of Energy by Lawrence Livermore National Laboratory under Contract No. DE-AC52-07NA27344
B.H.~would like to acknowledge support from the UC Laboratory Fees Research Program LFR-20-653698. K.J.R.~acknowledges support from NSF GRFP. LLNL-JRNL-825223.
\end{acknowledgments}

\bibliography{references,references_3}

%\bibliography{apssamp}% Produces the bibliography via BibTeX.

\end{document}